# Zero-inflated binary Tree Pólya splitting regression for multivariate count data

**Fabrice Moudjieu[1,2], Jean Peyhardi[3], Maxime Réjou-Méchain [4], Patrice Soh Takam[5],**

**Frédéric Mortier [4]**

[1] Ecole Nationale Supérieure Polytechnique de Yaoundé, Yaoundé, Cameroun.

[2] UPR Forêts et Sociétés, CIRAD, Montpellier, 34398, France.

[3] IMAG, CNRS, Université de Montpellier, Montpellier, 34090, France.

[4] AMAP, Université de Montpellier, CIRAD, CNRS, INRAE, IRD, Montpellier, France.

[5] Département de mathématiques, Université de Yaoundé 1, Yaoundé, Cameroun.

Summary: Species distribution models (SDMs) are widely used to assess the effects of environmental factors on species distributions. However, classical SDMs ignore inter-species dependencies. Multivariate SDMs (MSDMs), especially those based on latent Gaussian fields such as the multivariate Poisson log-normal (MPLN), address this limitation but face challenges related to computation, dimensionality, and interpretability. Pólya-splitting (PS) distributions offer an alternative, combining a model for total abundance with a multivariate allocation structure, and have natural interpretations from ecological process models. Yet, they lack flexibility in modeling correlation structures. Tree Pólya-splitting (TPS) distributions overcome this by introducing hierarchical structure such as a phylogenetic tree. In this paper, we extend TPS to account for zero-inflation, leading to the zero-inflated tree Pólya-splitting (Z-TPS) family. We detail its statistical properties, show how standard software enables efficient inference, and illustrate its ecological relevance using tree abundance data from over 180 genera across the Congo Basin tropical rainforest.

## 1. Introduction

In statistical ecology, species distribution models (SDMs) are widely used to infer the impacts of environmental factors on species distributions, both temporally and spatially (Elith and Leathwick, 2009). SDMs primarily combine the generalized linear model framework with various inference algorithms. Moreover, SDMs, which are species-by-species approaches, cannot capture key ecological mechanisms such as biotic interactions or functional or evolutionary relatedness.

To overcome these limitations, multivariate SDMs (MSDMs) have been developed to account for relationships between species (Ovaskainen et al., 2010). In ecology, MSDMs are mostly based on the assumption of a partially interpretable latent multivariate Gaussian random field (Warton et al., 2015). For example, multivariate individual random effects have been used to model species co-occurrence by characterizing correlation structures as a function of phylogenetic history (Ives and Helmus, 2011), or by exploiting the equivalence between link functions and latent variables, such as the logit or probit links with logistic or Gaussian distributions (O'Brien and Dunson, 2004; Chib and Greenberg, 1998). For count data, the multivariate Poisson log-normal (MPLN, Aitchison and Ho (1989)) distribution is currently advocated due to its flexibility (Ovaskainen et al., 2017). MPLN is a hierarchical (two-stage) model: at the first level, species abundances are conditionally independent given latent random processes; at the second level, the log-processes are assumed to follow a multivariate Gaussian distribution.

MSDMs have also been referred to as co-occurrence models (Ovaskainen et al., 2010), hierarchical Bayesian species distribution models (hSDMs, Clément and Vieilledent (2023)), hierarchical modelling of species communities (HMSC, Tikhonov et al. (2022)), or joint species distribution models (JSDMs, Pollock and et al. (2014)). In what follows, we use the terms 'joint' and 'multivariate' interchangeably. Although the latent Gaussian approach has emerged as a versatile framework for the analysis of joint species abundances, it presents several limitations.

The first is computational: exact inference is challenging due to the presence of high-dimensional latent variables (Inouye et al., 2017). Over the past decades, considerable efforts have been made to develop fast and stable inference algorithms using maximum likelihood (Gibaud et al., 2025), Markov Chain Monte Carlo (Tikhonov et al., 2022; Clément and Vieilledent, 2023),

Monte Carlo integration (Pichler and Hartig, 2021), Laplace approximations, or variational methods (Serhiyenko et al., 2016; Chiquet et al., 2019a). The second limitation lies in the high dimensionality of the parameter space, which grows quickly with the number of species: from roughly sixty parameters for $J = 10$ species to about six thousand for $J = 100$, and exceeding twenty thousand parameters for $J = 200$ species, a range that matches tree species richness commonly observed per hectare in tropical forests (Gentry, 1988). To address this, several strategies for dimension reduction and regularization have been proposed (Warton et al., 2015), primarily based on latent factor models (Bartholomew et al., 2011) or graphical lasso techniques (Friedman et al., 2007; Chiquet et al., 2019a). The third limitation is conceptual: inter-species dependencies are modeled indirectly through latent variables and may not reflect true ecological mechanisms. This can result in underestimating species relationships (Chiquet et al., 2019a). The fourth challenge involves modeling dependency structures as functions of environmental factors as species interactions likely vary depending on the ecological contexts.

Alternatively, in microbiome studies, regression models with a multinomial likelihood have been employed. Because observed read counts are strongly influenced by sequencing depth, genera abundances cannot be directly compared (Manor and Borenstein, 2015). Only relative abundances can be inferred, motivating the use of singular distributions, supported on the discrete simplex, such as the multinomial, or its more flexible generalizations: the Dirichlet-multinomial (DM) and generalized Dirichlet-multinomial distributions (Chen and Li, 2013; Wang and Zhao, 2017). This explains why such distributions are rarely used in modeling species distributions, where total or absolute abundances are meaningful.

Pólya-splitting (PS) distributions, recently introduced by Peyhardi et al. (2021), address these limitations. PS distributions extend the multivariate Pólya distribution (Dyczka, 1973) by modeling total species abundance with a univariate distribution and splitting this total among species using a multivariate Pólya distribution. This framework encompasses a wide family, including the multinomial, negative multinomial, multivariate hypergeometric, Dirichlet-multinomial, and generalized Waring distributions. More importantly, PS distributions emerge as natural stationary distributions of multivariate birth-death processes used in neutral theory (Peyhardi et al., 2024), allowing for process-based interpretations compared to standard correlational models like the MPLN. However, PS distributions cannot accommodate flexible correlation structures. The model imposes fixed correlation patterns, typically with the same sign for all pairwise correlations. To address this, Valiquette et al. (2024) introduced tree Pólya-splitting (TPS) distributions.

TPS models combine successive PS distributions along a known partition tree. TPS distributions offer several appealing statistical properties, particularly for computation: the log-likelihood is separable at each node of the tree, allowing inference to be parallelized using standard statistical software. All `R` code is based on existing packages (see Section 3.2). From an ecological perspective, choosing specific trees allows researchers to test biological hypotheses, such as disentangling the effects of evolutionary history from those of environmental filtering. However, TPS models do not account for the zero-inflation often observed in count data from biodiverse ecosystems.

Multivariate zero-inflation has been studied at the community level, where all species are either present or absent (Zhang et al., 2022), or at the species level, assuming that zero-inflation affects each species independently. For example, Batardière et al. (2024a) proposed an extension of PLN models where each species' abundance is modeled using a zero-inflated Poisson mixture, or alternatively, a single zero-inflation parameter is shared across all species. These approaches only capture a subset of plausible zero-inflation mechanisms. More recently, Tang and Chen (2019) proposed a novel zero-inflated version of the generalized Dirichlet multinomial (Z-GDM), using its sequential construction.

In this paper, we introduce a new family of probability distributions: the zero-inflated binary tree Pólya-splitting (Z-TPS). This family generalizes TPS distributions on any binary tree and includes the GDM as a special case. Section 3 presents the construction of the Z-TPS family, focusing on binary trees. We derive marginal and conditional properties, with particular attention to the factorial moments and the covariances. We also demonstrate how standard `R` packages can be used for efficient and reproducible inference. The second part introduces the regression extension. Finally, we present results from a real-world case study involving the abundance of more than 180 tree genera sampled across 1,571 plots spanning over 6 million hectares of tropical rainforest in the Congo Basin, and the tree used in this model is the phylogenetic tree associated with these genera. Our results are evaluated in comparison to alternative modeling strategies, including Poisson, negative binomial, and multivariate Poisson log-normal (MPLN), both with and without zero-inflation, under either full or diagonal covariance structures.

## 2. Tree Pólya-splitting distributions

This section introduces the notation, definitions and key properties of the recently proposed class of tree Pólya-splitting distributions for multivariate count data (Peyhardi and Fernique, 2017; Jones and Marchand, 2019; Peyhardi et al., 2021; Valiquette et al., 2024).

### 2.1 *Pólya distributions*

Let $J \in \mathbb{N}^*$ be the number of species (variables) and $\boldsymbol{Y} = (Y_1, Y_2, ..., Y_J)$ the random count vector of abundances of each species over $I$ samples plots. In the application, $J = 180$ and $I = 1,571$ sampled plots (of surface $10 \times 10$ km$^2$) from central African rainforests. We denote $|\boldsymbol{Y}| = \sum_{j=1}^{J} Y_j$ the sum of $\boldsymbol{Y}$ components; $\Delta_n = \{\boldsymbol{y} \in \mathbb{N}^J : |\boldsymbol{y}| = n\}$ the discrete simplex on $\mathbb{N}^J$; $\blacktriangle_n = \{\boldsymbol{y} \in \mathbb{N}^J : |\boldsymbol{y}| \leqslant n\}$ the discrete corner of the hypercube; $\blacksquare_{\boldsymbol{\theta}} = \{\boldsymbol{y} \in \mathbb{N}^J : y_1 \leqslant \theta_1, \ldots, y_J \leqslant \theta_J\}$ the hyper-rectangle;

$\theta_j \in \mathbb{R}^+$ and $\boldsymbol{\theta} = (\theta_1, \theta_2, \ldots, \theta_J)$. The random count vector $\boldsymbol{Y}$ follows the multivariate (singular) Pólya distribution $\mathcal{P}^{[c]}_{\Delta_n}(\boldsymbol{\theta})$ if for all $\boldsymbol{y} = (y_1, y_2, \ldots, y_J) \in \Delta_n$, its conditional probability mass function (pmf) is:

$$\mathbf{p}_{|\boldsymbol{Y}|=n}(\mathbf{y}) = \frac{n!}{(|\boldsymbol{\theta}|)_{(n,c)}} \prod_{j=1}^{J} \frac{(\theta_j)_{(y_j,c)}}{y_j!},$$

where $(\theta)_{(n,c)}$ corresponds to the *generalized factorial* given by

$$(\theta)_{(n,c)} \begin{cases} 1 & \text{if } n = 0; \\ \theta(\theta+c)\ldots(\theta+(n-1)c) & \text{if } n \geqslant 1. \end{cases}$$

and $c \in \{-1, 0, 1\}$. The non-singular version of the multivariate Pólya distribution is denoted by $\mathcal{P}^{[c]}_{\blacktriangle_n}(\boldsymbol{\theta}, \gamma)$ and is supported on $\blacktriangle_n$. The univariate version ($J = 1$), necessary non-singular, is denoted by $\mathcal{P}^{[c]}_n(\theta, \gamma)$ and is supported on $\{0, \ldots, n\}$ (e.g., the binomial distribution).

## 2.2 *Pólya-splitting distributions*

Multivariate Pólya distributions include three well-known multivariate distributions: the multivariate hypergeometric ($c = -1$), the multinomial ($c = 0$) and the Dirichlet multinomial ($c = 1$) distributions (Dyczka, 1973). However, since multivariate Pólya distributions are supported on the discrete simplexe, they are not suitable for modeling the variability of all the components of random count vectors. Indeed, any component of $(Y_1, \ldots, Y_J)$ is deterministic when the $J-1$ other components are known. To overcome this limitation, Peyhardi et al. (2021) extended the multivariate Pólya distributions into a more general family, named the Pólya-splitting distribution (PSD). This family consists in assuming the total abundance, $n$, as a random variable. Peyhardi et al.'s original and powerful results were to highlight how the choice of total abundance distribution impacts the structure of dependencies within the random count vector. The formalism of this subsection is based on (Peyhardi et al., 2021) and (Peyhardi et al., 2024).

The random count vector $\boldsymbol{Y}$ follows a Pólya-splitting distribution if the total random count variable $|\boldsymbol{Y}|$ follows a univariate count distribution $\mathcal{L}(\psi)$ and conditionally to the event $|\boldsymbol{Y}| = n$ the vector $\boldsymbol{Y}$ follows the multivariate Pólya distribution $\mathcal{P}^{[c]}_{\Delta_n}(\boldsymbol{\theta})$. This distribution will be denoted by

$$\mathcal{P}^{[c]}_{\Delta_n}(\boldsymbol{\theta}) \underset{n}{\wedge} \mathcal{L}(\psi),$$

where $\underset{n}{\wedge}$ denotes the mixing operator, indicating that the parameter $n$ is random and follows the distribution $\mathcal{L}(\psi)$. In the context of joint species distribution model (JSDM), the distribution $\mathcal{L}(\psi)$ will be named the *global abundance distribution* and the Pólya distribution will be named the *split distribution*. The pmf of such a Pólya-splitting distribution is given by

$$\mathbf{p}(\mathbf{y}) = \mathbf{p}(|\mathbf{y}|) \frac{|\mathbf{y}|!}{(|\boldsymbol{\theta}|)_{(|\mathbf{y}|,c)}} \prod_{j=1}^{J} \frac{(\theta_j)_{(y_j,c)}}{y_j!},$$

for all $\boldsymbol{y} = (y_1, y_2, \ldots, y_J) \in \mathbb{N}^J$. The $j^{\text{th}}$-univariate marginal of the Pólya-splitting distributions is the damage process

$$\mathcal{P}^{[c]}_n(\theta_j, |\boldsymbol{\theta}_{-j}|) \underset{n}{\wedge} \mathcal{L}(\psi),$$

where $\boldsymbol{\theta}_{-j}$ denotes the vector of parameters $\boldsymbol{\theta}$ private of the $j^{\text{th}}$ component. The covariance of a pair $(Y_i, Y_j)$ is given by

$$\text{Cov}(Y_i, Y_j) = \frac{\theta_i \theta_j}{|\boldsymbol{\theta}|^2 (|\boldsymbol{\theta}| + c)} \left[ \mu_2 - (|\boldsymbol{\theta}| + c) \mu_1^2 \right] \tag{1}$$

where $\mu_k$ denotes the factorial moment of the global abundance distribution. Equation (1) shows that sign of covariances between species are constant whatever the pairs $(i, j)$. Moreover, for a given Pólya-splitting distribution, the sign of covariance is driven by the dispersion of the global abundance distribution. For multinomial-splitting distributions for instance ($c = 0$), the covariance of any pair is negative (resp. null or positive) when the global abundance is under (resp. equi or over) dispersed.

We present nine Pólya-splitting (PS) distributions in Table 1, each verifying a number of useful properties. First, they naturally arise in the context of the neutral theory of biodiversity—that is, they are the stationary distributions of multivariate birth–death processes under simple (neutral) assumptions on transition rates (Peyhardi et al., 2024). This generalizes the approach of Haegeman and Etienne (2008), who relaxed the zero-sum assumption (i.e., that total abundance remains constant over time) in the framework of relative species abundance in ecological communities, as popularized by Hubbell (2001). Until recently, the neutrality assumption (i.e., absence of species interactions) and the assumption of independence between species abundances were often conflated. Haegeman and Etienne (2017) proposed replacing the zero-sum assumption of Hubbell (2001) with an independence assumption. However, Peyhardi et al. (2024) demonstrated that neither of these assumptions is necessary to define a neutral model. Table 1 presents six neutral models (first and third rows) that relax the zero-sum assumption without assuming independence between species. This implies that correlations between species abundances may emerge even in the absence of biological interactions. Pólya-splitting distributions thus offer advantages in both modeling flexibility and ecological interpretation. Nevertheless, enforcing the same sign of correlation across all species pairs becomes unrealistic when the community size $J$ is large. To address this limitation, Valiquette et al. (2024) introduced the class of tree Pólya-splitting distributions, which recursively partition total abundance along a species tree.

| Split / Sum | Hypergeometric | Multinomial | Dirichlet multinomial | Sign of covariances |
|---|---|---|---|---|
| Pólya (1) | $\mathcal{H}_{\Delta_n}(\boldsymbol{\theta}) \overset{\wedge}{_n} \mathcal{H}_m(|\boldsymbol{\theta}|, \gamma)$ $=$ $\mathcal{H}_{\blacktriangle_m}(\boldsymbol{\theta}, \gamma)$ $\boldsymbol{\theta} \in \mathbb{N}^{*J},\ \gamma \in \mathbb{N}^*,\ m \in \mathbb{N}^*,\ m \leqslant |\boldsymbol{\theta}| + \gamma$ (support $= (\blacktriangle_m \setminus \blacktriangle_{m-\gamma}) \cap \blacksquare_{\boldsymbol{\theta}}$) | $\mathcal{M}_{\Delta_n}(\boldsymbol{\pi}) \overset{\wedge}{_n} \mathcal{B}_m(p)$ $=$ $\mathcal{M}_{\blacktriangle_m}(p \cdot \boldsymbol{\pi})$ $\boldsymbol{\pi} \in \Delta,\ m \in \mathbb{N}^*$ (support $= \blacktriangle_m$) | $\mathcal{DM}_{\Delta_n}(\boldsymbol{\theta}) \overset{\wedge}{_n} \beta\mathcal{B}_m(|\boldsymbol{\theta}|, \gamma)$ $=$ $\mathcal{DM}_{\blacktriangle_m}(\boldsymbol{\theta}, \gamma)$ $\boldsymbol{\theta} \in \mathbb{R}_+^{*J},\ \gamma \in \mathbb{R}_+^*,\ m \in \mathbb{N}^*$ (support $= \blacktriangle_m$) | Negatives |
| Power series (2) | $\mathcal{H}_{\Delta_n}(\boldsymbol{\theta}) \overset{\wedge}{_n} \mathcal{B}_{|\boldsymbol{\theta}|}(p)$ $=$ $\bigotimes_{j=1}^{J} \mathcal{B}_{\theta_j}(p)$ $\boldsymbol{\theta} \in \mathbb{N}^{*J},\ p \in (0,1)$ (support $= \blacksquare_{\boldsymbol{\theta}}$) | $\mathcal{M}_{\Delta_n}(\boldsymbol{\pi}) \overset{\wedge}{_n} \mathcal{P}(\lambda)$ $=$ $\bigotimes_{j=1}^{J} \mathcal{P}(\pi_j \lambda)$ $\boldsymbol{\pi} \in \Delta,\ \lambda \in \mathbb{R}_+^*$ (support $= \mathbb{N}^J$) | $\mathcal{DM}_{\Delta_n}(\boldsymbol{\theta}) \overset{\wedge}{_n} \mathcal{NB}(|\boldsymbol{\theta}|, p)$ $=$ $\bigotimes_{j=1}^{J} \mathcal{NB}(\theta_j, p)$ $\boldsymbol{\theta} \in \mathbb{R}_+^{*J},\ p \in (0,1)$ (support $= \mathbb{N}^J$) | Null |
| Inverse Pólya (3) | $\mathcal{H}_{\Delta_n}(\boldsymbol{\theta}) \overset{\wedge}{_n} \beta\mathcal{B}_{|\boldsymbol{\theta}|}(a,b)$ $=$ $\mathcal{M}\beta\mathcal{B}_{\blacksquare_{\boldsymbol{\theta}}}(a,b)$ $\boldsymbol{\theta} \in \mathbb{N}^{*J},\ a \in \mathbb{R}_+^*,\ b \in \mathbb{R}_+^*$ (support $= \blacksquare_{\boldsymbol{\theta}}$) | $\mathcal{M}_{\Delta_n}(\boldsymbol{\pi}) \overset{\wedge}{_n} \mathcal{NB}(a,p)$ $=$ $\mathcal{NM}(a, p \cdot \boldsymbol{\pi})$ $\boldsymbol{\pi} \in \Delta,\ a \in \mathbb{R}_+^*,\ p \in \mathbb{R}_+^*$ (support $= \mathbb{N}^J$) | $\mathcal{DM}_{\Delta_n}(\boldsymbol{\theta}) \overset{\wedge}{_n} \beta\mathcal{NB}(|\boldsymbol{\theta}|, a, b)$ $=$ $\mathrm{MGWD}(b, \boldsymbol{\theta}, a)$ $\boldsymbol{\theta} \in \mathbb{R}_+^{*J},\ a \in \mathbb{R}_+^*,\ b \in \mathbb{R}_+^*$ (support $= \mathbb{N}^J$) | Positives |

Table 1: Nine remarkable Pólya-splitting distributions. Rows correspond to total abundance distribution, columns the splitting distribution.

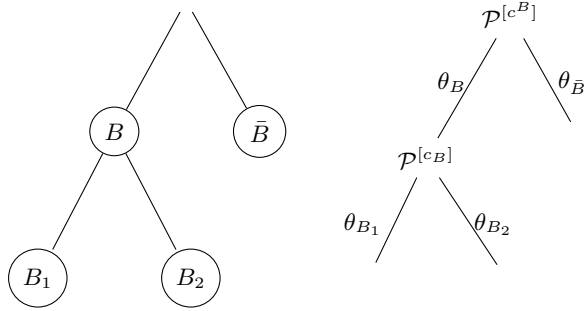

**Figure 1**: Notation of sibling and children nodes of a given internal node $B$ and their associated parameters.

### 2.3 *Binary Tree Pólya-splitting distribution*

The formalism of this subsection is based on (Valiquette et al., 2024) and is adapted to the specific case of binary partition tree. It is assumed that the definition of a directed tree is known; the reader may refer to Gross et al. (2018) for details. Let $\mathfrak{T} = (\mathcal{N}, \mathcal{E})$ denote a directed tree, where $\mathcal{N}$ is the set of nodes and $\mathcal{E}$ is the set of directed edges. Remark that $\mathcal{N}$ could be divided into the set of internal nodes $\mathcal{I}$ (i.e., nodes that have children) and the set of leaves $\mathcal{L}$ (i.e., nodes that have no children): $\mathcal{N} = \mathcal{I} \cup \mathcal{L}$ (disjoint). The root, denoted by $\Omega$, is the internal node that recursively gives the directions of all edges. In this paper, focus is made on the specific class of partition tree, i.e., such that

- the root is the set of all labels species: $\Omega = \{1, \ldots, J\}$,
- each leave is a singleton corresponding to one label species: $\mathcal{L} = \{\{1\}, \ldots, \{J\}\}$,
- sibling nodes form a non-trivial partition of their parent node.

Moreover, we reduce the study to the class of binary partition trees, i.e., such that each internal node has exactly two children. For a given internal node $B \in \mathcal{I}$, we will denote by $\bar{B}$ its unique sibling node and by $B_1$, $B_2$ its children nodes (see Figure 1 on the left). With these notations, we have $B = B_1 \cup B_2$ (disjoint) where $B_1$ and $B_2$ are non-empty subsets of $\Omega = \{1, \ldots, J\}$.

A binary tree splitting distribution is a multivariate count distribution that recursively split the total abundance $|\boldsymbol{y}_\Omega| = \sum_{j=1}^{J} y_j$ into the $J$ species, along the partition tree. For each internal node $B \in \mathcal{I}$, is associated the multivariate abundance $\boldsymbol{y}_B = \{y_j\}_{j \in B}$ of the species group $B$ and for each leave $\{j\} \in \mathcal{L}$, is associated the species abundance $y_j$. At each internal node $B \in \mathcal{I}$, the multivariate abundance $\boldsymbol{y}_B$ is factorized as follows:

$$p(\boldsymbol{y}_B) = p_{|\boldsymbol{y}_B|}(\boldsymbol{y}_{B_1}, \boldsymbol{y}_{B_2}) p(|\boldsymbol{y}_B|).$$

and thus we recursively obtain the factorization of the joint pmf

$$p(y_1, \ldots, y_J) = p(\boldsymbol{y}_\Omega) = \left\{ \prod_{B \in \mathcal{I}} p_{|\boldsymbol{y}_B|}(|\boldsymbol{y}_{B_1}|, |\boldsymbol{y}_{B_2}|) \right\} p(|\boldsymbol{y}_\Omega|). \tag{2}$$

This factorization is always true, and the model assumptions consist in assuming a parametric form for the pmf $p(|\boldsymbol{y}_\Omega|)$ and for the bivariate conditional pmf $p_{|\boldsymbol{y}_B|}(\boldsymbol{y}_{B_1}, \boldsymbol{y}_{B_2})$ at each internal node $B \in \mathcal{I}$. In the following, the parametric form of the global abundance distribution is not specified, but the reader may keep in mind a Poisson or a negative binomial distribution to fix the idea. Then, the split distribution is assumed to be a bivariate Pólya distribution. More formally, we have the following definition.

DEFINITION 1: Let $\mathfrak{T} = (\Omega, \mathcal{I}, \mathcal{L})$ denote a partition tree with $\boldsymbol{c} = \{c_B\}_{B \in \mathcal{I}}$ and $\boldsymbol{\theta} = \{(\theta_{B_1}, \theta_{B_2})\}_{B \in \mathcal{I}}$ two sets of parameters such that $c_B \in \{-1, 0, 1\}$ and $(\theta_{B_1}, \theta_{B_2}) \in \mathbb{R}_+^2$. The random count vector $\boldsymbol{Y} = \{Y_j\}_{j \in \Omega}$ is said to follow a tree Pólya-splitting distribution, denoted by $\mathcal{TP}_{\Delta_n}^{[\boldsymbol{c}]}(\boldsymbol{\theta}) \underset{n}{\wedge} \mathcal{L}(\psi)$, if

- the global abundance is distributed as $|\boldsymbol{Y}| \sim \mathcal{L}(\psi)$,
- for each internal node $B \in \mathcal{I}$, given $|\boldsymbol{y}_B| = n$ we have $|\boldsymbol{Y}_{B_1}| \sim \mathcal{P}_n^{[c_B]}(\theta_{B_1}, \theta_{B_2})$.

The joint pmf of such a distribution is given by

$$p(y_1, \ldots, y_J) = \left\{ \prod_{B \in \mathcal{I}} \binom{|\boldsymbol{y}_B|}{|\boldsymbol{y}_{B_1}|} \frac{(\theta_{B_1})_{(|\boldsymbol{y}_{B_1}|; c_B)} (\theta_{B_2})_{(|\boldsymbol{y}_{B_2}|; c_B)}}{(\theta_{B_1} + \theta_{B_2})_{(|\boldsymbol{y}_B|; c_B)}} \right\} p(|\boldsymbol{y}_\Omega|) \tag{3}$$

To describe the marginal and the moments we first need to introduce the notation of ancestors. For a given label species $j$, let $\mathcal{A}_j$ denote the set of all ancestors of $\{j\}$ (i.e., nodes that are parent of $j$ with any order) except the root and including $\{j\}$ himself. The marginal distribution of $Y_j$ is given by the recursive damage process

$$Y_j \sim \left\{ \bigwedge_{n_B : B \in \mathcal{A}_j} \mathcal{P}_{n_B}^{[c^B]}(\theta_B, \theta_{\bar{B}}) \right\} \wedge \mathcal{L}(\psi), \tag{4}$$

where $c^B$ denotes the parameter $c$ associated with the parent node of $B$ (not to be confused with $c_B$). The marginal pmf is thus given by

$$p(y_j) = \left\{ \prod_{B \in \mathcal{A}_j} \binom{|\boldsymbol{y}_B| + |\boldsymbol{y}_{\bar{B}}|}{|\boldsymbol{y}_B|} \frac{(\theta_B)_{(|\boldsymbol{y}_B|; c^B)} (\theta_{\bar{B}})_{(|\boldsymbol{y}_{\bar{B}}|; c^B)}}{(\theta_B + \theta_{\bar{B}})_{(|\boldsymbol{y}_B| + |\boldsymbol{y}_{\bar{B}}|; c^B)}} \right\} \times p(|\boldsymbol{y}_\Omega|)$$

The expectation of species abundance is given by

$$\mu(j) = \mathrm{E}[Y_j] = \left\{ \prod_{B \in \mathcal{A}_j} p_B \right\} \times \mu(\Omega), \tag{5}$$

where $p_B = \frac{\theta_B}{\theta_B + \theta_{\bar{B}}}$ is the mean proportion of species group $B$ versus $\bar{B}$ and $\mu(\Omega)$ is the expectation of the global abundance. The mean abundance of species $j$ is thus the global mean abundance, successively damaged by the mean proportion along their memberships groups. According to Valiquette et al. (2024), the factorial moment of order two is given by

$$\mu_2(j) = \mathrm{E}[Y_j(Y_j - 1)] = \left\{ \prod_{B \in \mathcal{A}_j} p_B \cdot p_B^{[c^B]} \right\} \times \mu_2(\Omega), \tag{6}$$

where $p_B^{[c^B]} = \frac{\theta_B + c^B}{\theta_B + \theta_{\bar{B}} + c^B}$. To describe the covariance between species abundances $Y_i$ and $Y_j$ we need to introduce the notation $S = S(i, j)$ as the separator node between leaves $\{i\}$ and $\{j\}$, i.e., the younger common ancestor of leaves $\{i\}$ and $\{j\}$ or equivalently the unique leaf of the chain $\mathcal{A}_i \cap \mathcal{A}_j$ (intersection between ancestors of $\{i\}$ and $\{j\}$). This separator node $S$ has necessary two children $S_i$ and $S_j$ that are ancestors of $\{i\}$ and $\{j\}$ respectively. The covariance is given by

$$\mathrm{Cov}(Y_i, Y_j) = \left\{ \prod_{B \in \mathcal{A}_i^j} p_B \right\} \times \left\{ \prod_{B' \in \mathcal{A}_j^i} p_{B'} \right\} \times \mathrm{Cov}(|\boldsymbol{Y}_{S_i}|, |\boldsymbol{Y}_{S_j}|),$$

where $A_i^j := \mathcal{A}_i \setminus \mathcal{A}_i \cap \mathcal{A}_j$ and the covariance between abundances of species groups $S_i$ and $S_j$ is given by the formula of a simple Pólya-splitting distribution

$$\mathrm{Cov}(|\boldsymbol{Y}_{S_i}|, |\boldsymbol{Y}_{S_j}|) = p_{S_i} \cdot p_{S_j} \frac{\theta_{S_i} + \theta_{S_j}}{\theta_{S_i} + \theta_{S_j} + c_S} \mu_2(S) - p_{S_i} \cdot p_{S_j} \cdot \mu^2(S)$$

and $\mu_k(S)$ is the factorial moment of group species abundances $z_S$ (for $k = 1, 2$) with similar formula (5) and (6) as $\mu_k(j)$ along the path $\mathcal{A}_i \cap \mathcal{A}_j$ instead of $\mathcal{A}_j$.

## 3. Zero inflation and regression framework

### 3.1 *Zero inflation*

The global abundance may simply be zero-inflated since this is a univariate count distribution. Let $\pi \in [0, 1]$ be the proportion of zero for the global abundance, then zero-inflated distribution is given by

$$\pi \delta_0 + (1 - \pi) \mathcal{L}(\psi),$$

where $\delta_0$ denotes the Dirac distribution at point 0. The $k^{th}$ factorial moment of this zero-inflated global abundance distribution is therefore $\tilde{\mu}_k(\Omega) = (1 - \pi)\mu_k(\Omega)$ for $k \in \mathbb{N}^*$ where $\mu_k(\Omega)$ denotes the $k^{th}$ factorial moment associated with the distribution $\mathcal{L}(\psi)$. Notice that the expectation $\mu_1(\Omega)$ is simply denoted by $\mu(\Omega)$. It should be kept in mind that a zero on the total abundances is equivalent to a multivariate zero, i.e., all species have jointly no occurrence. This situation is extremely rare, especially when the number of species $J$ is important.

It is then more appropriated to define a zero inflation on the different binary splits, meaning that some group of species $B_1$ and/or $B_2$ may have no occurrence at the same site. At each internal node $B \in \mathcal{I}$ we assume a bivariate zero-inflated Pólya distribution for the couple of groups abundances $(|\boldsymbol{Y}_{B_1}|, |\boldsymbol{Y}_{B_2}|)$, knowing the total abundance of both groups $|\boldsymbol{Y}_B| = n$, as follows:

$$\pi_{B_1} \delta_{(0,n)} + \pi_{B_2} \delta_{(n,0)} + (1 - \pi_{B_1} - \pi_{B_2}) \mathcal{P}_n^{[c_B]}(\theta_{B_1}, \theta_{B_2}).$$

This distribution is well-defined if $\pi_{B_1} + \pi_{B_2} \in [0, 1]$. Using similar argument than for the global abundance, the proportion of zero for a given group of species $B$ is expected to be small or null for a big group (closed to the root) and bigger for a small group (closed to the leaves). The advantage of this model compared with the zero-inflated GDM is that groups have an interpretation in terms of phylogeny contrarily to a sequential tree that forms groups without information. Formally we have

DEFINITION 2: Let $\mathfrak{T} = (\Omega, \mathcal{I}, \mathcal{L})$ denote a partition tree with $\boldsymbol{c} = \{c_B\}_{B \in \mathcal{I}}$, $\boldsymbol{\pi} = \{(\pi_{B_1}, \pi_{B_2})\}_{B \in \mathcal{I}}$ and $\boldsymbol{\theta} = \{(\theta_{B_1}, \theta_{B_2})\}_{B \in \mathcal{I}}$ three sets of parameters such that $c_B \in \{-1, 0, 1\}$, $(\pi_{B_1}, \pi_{B_2}) \in [0, 1]^2$, $\pi_{B_1} + \pi_{B_2} \in [0, 1]$, $(\theta_{B_1}, \theta_{B_2}) \in \mathbb{R}_+^2$. The random count vector $\boldsymbol{Y} = \{Y_j\}_{j \in \Omega}$ is said to follow a zero-inflated tree Pólya-splitting distribution, denoted by $\mathcal{ZTP}_{\Delta_n}^{[c]}(\boldsymbol{\pi}; \boldsymbol{\theta}) \underset{n}{\wedge} \mathcal{L}(\psi)$, if

- the global abundance is distributed as $|\boldsymbol{Y}| \sim \mathcal{L}(\psi)$,
- for each internal node $B \in \mathcal{I}$, given $|\boldsymbol{y}_B| = n$ we have $|\boldsymbol{Y}_{B_1}| \sim \pi_{B_1} \delta_{(0,n)} + \pi_{B_2} \delta_{(n,0)} + (1 - \pi_{B_1} - \pi_{B_2}) \mathcal{P}_n^{[c_B]}(\theta_{B_1}, \theta_{B_2})$.

The joint and marginal pmfs of such a zero-inflated binary tree Pólya-splitting distribution are similar to (3) and (4) where the conditional pmf of the binary split at node $B$ is given by

$$p_{|\boldsymbol{y}_B|}(|\boldsymbol{y}_{B_1}|, |\boldsymbol{y}_{B_2}|) = \pi_{B_1} \mathbb{1}_{|\boldsymbol{y}_{B_1}| = 0} + \pi_{B_2} \mathbb{1}_{|\boldsymbol{y}_{B_2}| = 0} + (1 - \pi_{B_1} - \pi_{B_2}) \binom{|\boldsymbol{y}_B|}{|\boldsymbol{y}_{B_1}|} \frac{(\theta_{B_1})_{(|\boldsymbol{y}_{B_1}|; c_B)} (\theta_{B_2})_{(|\boldsymbol{y}_{B_2}|; c_B)}}{(\theta_{B_1} + \theta_{B_2})_{(|\boldsymbol{y}_B|; c_B)}}.$$

As in the previous subsection, the expectation of species abundance is given by the formula

$$\mu(j) = \mathrm{E}[Y_j] = \left\{ \prod_{B \in \mathcal{A}_j} \tilde{p}_B \right\} \times \mu(\Omega), \tag{7}$$

where $\tilde{p}_B = \pi_B + (1 - \pi_B - \pi_{\bar{B}}) \dfrac{\theta_B}{\theta_B + \theta_{\bar{B}}}$ denotes the mean proportion of species group $B$ under a zero-inflated Pólya-splitting distribution. More generally, the $k^{th}$ factorial moment of the abundance of the $j^{th}$ species is given by (details in Appendix A1)

$$\mu_k(j) = \mathrm{E}[Y_j(Y_j - 1) \dots (Y_j - k + 1)] = \left\{ \prod_{B \in \mathcal{A}_j} \tilde{p}_B^{(k)} \right\} \times \mu_k(\Omega), \quad \text{where} \quad \tilde{p}_B^{(k)} = \pi_B + (1 - \pi_B - \pi_{\bar{B}}) \frac{\prod_{t=0}^{k-1}(\theta_B + c^B t)}{\prod_{t=0}^{k-1}(\theta_B + \theta_{\bar{B}} + c^B t)}. \tag{8}$$

Furthermore, it can be shown that, for a given species $j$, the marginal distribution of $Y_j$ is a zero-inflated distribution (details in Appendix A2):

$$p_j \delta_0 + (1 - p_j) \mathcal{L}_j,$$

where the proportion of zero-inflation is given by

$$p_j = 1 - \prod_{B \in \mathcal{A}_j} (1 - \pi_B),$$

and $\mathcal{L}_j$ is a mixture of some damage Pólya distributions. The proportion of zero-inflation for a species $j$ is therefore a consequence of zero-inflation proportions of its group's ancestors.

Finally, the covariance is given by (details in Appendix A3)

$$\text{Cov}(Y_i, Y_j) = \left\{ \prod_{B \in \mathcal{A}_i^j} \tilde{p}_B \right\} \times \left\{ \prod_{B' \in \mathcal{A}_j^i} \tilde{p}_{B'} \right\} \times \text{Cov}(|\boldsymbol{Y}_{S_i}|, |\boldsymbol{Y}_{S_j}|),$$

where the covariance between abundances of species groups $S_i$ and $S_j$ is given by the formula of a zero-inflated Pólya-splitting distribution

$$\text{Cov}(|\boldsymbol{Y}_{S_i}|, |\boldsymbol{Y}_{S_j}|) = (1 - \pi_S) \cdot p_{S_i} \cdot p_{S_j} \frac{\theta_{S_i} + \theta_{S_j}}{\theta_{S_i} + \theta_{S_j} + c_S} \mu_2(S) - \tilde{p}_{S_i} \cdot \tilde{p}_{S_j} \cdot \mu^2(S)$$

where $\pi_S = \pi_{S_i} + \pi_{S_j}$ and $\mu_k(S)$ is the $k^{th}$ factorial moment of group species abundances $z_S$ with similar formula (8) as $\mu_k(j)$ along the path $\mathcal{A}_i \cap \mathcal{A}_j$ instead of $\mathcal{A}_j$.

### 3.2 *Regression framework*

Consider a regression framework and denote by $x = (1, x_1, \ldots, x_p) \in \mathbb{R}^{p+1}$ the vector of explanatory variables. In the context of a splitting model, the regression is performed separately on the sum and the split (Peyhardi et al., 2021). Following this idea with a binary splitting model, the regression is performed separately on the sum and successively on each binary split. For the sum, the link function

$$g\{\mu(\Omega)\} = x^T \beta^\Omega$$

could be the log function in a case of a Poisson distribution $\mathcal{P}(\lambda)$ or a logit function on the $p$ parameter of a negative binomial distribution $\mathcal{NB}(r, p)$. The vector of parameters $\beta^\Omega$ gives informations about the explanatory variables effects on the global abundance. Then, at each internal node $B \in \mathcal{I}$, the proportion of species group $p_{B_1} = \theta_{B_1}/(\theta_{B_1} + \theta_{B_2})$ (among both groups $B = B_1 \cup B_2$) is linked to a linear predictor through a logit link function

$$\text{logit}(p_{B_1}) = x^T \beta^{B_1} \quad \Leftrightarrow \quad p_{B_1} = F(x^T \beta^{B_1}),$$

where $F$ denotes the logistic cumulative distribution function (cdf). Notice that the proportion of the other group $B_2$, obtained as the complementary, is also linked with logit function

$$p_{B_2} = 1 - F(x^T \beta^{B_1}) \quad \Leftrightarrow \quad \text{logit}(p_{B_2}) = x^T \beta^{B_2},$$

with $\beta^{B_2} = -\beta^{B_1}$. Another absolutely continuous cdf could also be used, with symmetry property (i.e., $1 - F(\eta) = F(-\eta)$) in order to obtain equivalent model whenever the choice of the reference group $B_1$ or $B_2$. We will denote by $\beta^B$ the parameter $\beta^{B_1}$ or its opposite according to the reference choice. Moreover, a second parameter $\sigma^B = 1/(\theta_{B_1} + \theta_{B_2})$ is linked to a predictor through a log link function

$$\log(\sigma^B) = x^T \delta^B.$$

It should be noted that assume a logit link on $p_{B_1}$ and a log link on $\sigma_B$ is equivalent to the regression of both parameters $\theta_{B_1}$ and $\theta_{B_2}$ through log link functions[1].

In the more general form, the proportions of zero-inflation could also be linked to the explanatory variables in the same way

$$\pi_{B_1} = F(x^T b^{B_1}) \quad \text{and} \quad \pi_{B_2} = F(x^T b^{B_2}).$$

Note that the proportion $p_{B_1} = \theta_{B_1}/(\theta_{B_1} + \theta_{B_2})$ cannot be considered as continuous in the case $c = -1$ of an hypergeometric split distribution. The regression framework has not been defined in this case. Focus will be thus made on multinomial ($c = 0$) and Dirichlet multinomial ($c = 1$) cases in the following. Remark that in the case $c = 0$ the parameter $\sigma^B = 1/(\theta_{B_1} + \theta_{B_2})$ is not identifiable and therefore only the proportion $p_{B_1}$ is regressed.

*Parameters inference.* Let $\Theta = \{\beta^\Omega, (\beta^B, \delta^B)_{B \in \mathcal{I}}, (b^{B_1}, b^{B_2})_{B \in \mathcal{I}}\}$ the set of all regression parameters. Following the factorization (2) of the joint pmf, the log-likelihood is decomposed into the log-likelihood of the global abundance and those of split at each internal node

$$l(y_1, \ldots, y_J; \Theta) = l(|\boldsymbol{y}_\Omega|; \beta^\Omega) + \sum_{B \in \mathcal{I}} l\{|\boldsymbol{y}_{B_1}|, |\boldsymbol{y}_{B_2}|; (\beta^B, \delta^B, b^{B_1}, b^{B_2})_{B \in \mathcal{I}}\}. \tag{9}$$

Since there is no assumption of contrast between parameters of different internal node, the log-likelihood can be separately maximized. The left part corresponds to the inference of a generalized linear model (GLM) for a univariate count response based on total abundances dataset $\{(|\boldsymbol{y}_i|, x_i)_{i=1,\ldots,I}\}$. The log-likelihood formula (9) has been written for only one observation, but it would be obviously be summed over all observations $i = 1, \ldots, I$. We choose a negative binomial response distribution to take into account the sur-dispersion of the global abundance. It is possible to compare different models (e.g. Poisson, negative binomial, etc $\ldots$) using a penalized criteria such as the Akaike information criterion (AIC) that is decomposed along the internal nodes as the log-likelihood. Then, the parameters of the zero-inflated Pólya distribution are separately estimated at each internal node $B \in \mathcal{I}$. The parameter $c_B$ is selected among $\{0, 1\}$, i.e., the multinomial and Dirichlet multinomial split are compared using AIC. In our case, the Dirichlet multinomial is always preferred. Then, for the inference of parameters $\theta_{B_1}, \theta_{B_2}$ and $\pi_{B_1}, \pi_{B_2}$, it is remarked that if at least one of both zero-inflation parameters is null, then the bivariate split model can be viewed as a univariate zero-inflated binomial ($c = 0$) or beta binomial ($c = 1$) regression model, whose inference is already known (Rigby et al., 2019). Therefore, for the binomial and beta binomial cases, we compared, using the AIC, three kinds of regression models: the (beta) binomial on $|\boldsymbol{y}_{B_1}|$ (case $\pi_{B_1} = \pi_{B_2} = 0$), the zero-inflated (beta) binomial on $|\boldsymbol{y}_{B_1}|$ (case $\pi_{B_1} \neq 0, \pi_{B_2} = 0$) and the zero-inflated (beta) binomial on $|\boldsymbol{y}_{B_2}|$ (case $\pi_{B_1} = 0, \pi_{B_2} \neq 0$).

---

[1] Both parametrizations could be used according to the selected software.

*Parameters interpretation.* Since, the mean abundance of each species $j$ is a product of form (5), the effect of a given explanatory variable $x_k$ on $\mu(j)$ can not directly be interpreted regarding the parameters $\{\beta^B\}_{B \in \mathcal{A}_j}$ along the ancestors of leave $\{j\}$. Therefore, we propose to use the size effect of an explanatory variable $x_k$ on the species abundance mean $\mu(j)$. Using recursively the derivative of a product, it is obtained that

$$\frac{\partial \mu(j)}{\partial x_k} = \left\{ \sum_{B \in \mathcal{A}_j} \frac{\partial \tilde{p}_B}{\partial x_k} \prod_{\substack{B' \in \mathcal{A}_j \\ B' \neq B}} \tilde{p}_{B'} \right\} \mu(\Omega) + \left\{ \prod_{B \in \mathcal{A}_j} \tilde{p}_B \right\} \frac{\partial \mu(\Omega)}{\partial x_k}.$$

Note that if $\mu(\Omega) = \exp(x^T \beta^\Omega)$ then the right part of the sum is equal to $\beta_k^\Omega \cdot \mu(j)$. For the left part, recall that $\bar{p}_B = \pi_{\bar{B}} + (1 - \pi_B - \pi_{\bar{B}}) p_B$ is the mean proportion of species group $B$ under a zero-inflated Pólya-splitting regression and $\frac{\partial p_B}{\partial x_k} = \beta_k^B \cdot f(x^T \beta^B)$ where $f = F'$ denotes the logistic probability density function (pdf).

We have seen that the effect of an explanatory variable $x_k$ on a leave $\{j\}$ is computed along the path of its ancestors. Whenever, other parameters interpretations are available within a tree Pólya-splitting model. Firstly, the vector of parameters $\beta^\Omega$ is simply interpreted as effects on the global mean abundance, following the GLM framework for negative binomial response. Secondly, for each internal node $B \in \mathcal{I}$, the vectors of parameters $(\beta^B, \delta^B)$ are interpreted as effects on mean proportions of groups $B_1$ and $B_2$ whereas the vectors of parameters $(b^{B_1}, b^{B_2})$ are interpreted as effect on absence of groups $B_1$ and $B_2$, following the GLM framework for a zero-inflated (beta) binomial response. It is then possible to separately study the environmental factors that influence the global abundance of a community and those distinguishing phylogenetic groups of species.

## 4. Application

This application aims to illustrate and discuss the performance of the proposed models compared to simplified or alternative approaches. Rather than relying on simulated data, we focus on the large *CoFor* dataset (Réjou-Méchain et al., 2021) to evaluate model behavior under realistic ecological conditions. Although simulation studies provide a controlled framework for investigating specific mechanisms, empirical data from tropical forests naturally combine high species richness, pronounced environmental heterogeneity, and pervasive sparsity—features that are difficult to reproduce jointly in simulation-based settings. Model comparisons are therefore conducted using a range of complementary criteria, encompassing both statistical performance (goodness-of-fit and predictive accuracy) and ecological interpretability.

### 4.1 *CoFor Data*

The CoFor data consist of counts of $J = 180$ tree genera monitored across $I = 1571$ plots (statistical units), covering more than 90,000 hectares of Central African rainforests. Each plot aggregates a variable number of 0.5-hectare subplots, with this information used as offsets in the analysis. Accordingly, 24 climatic variables are available at the Central African scale (Réjou-Méchain et al., 2021). For this study, only the first three principal components of these variables are used as environmental factors. The first principal component corresponds to climatic conditions contrasting areas with a cool, light-deficient dry season (coastal Gabon) and areas with high evapotranspiration rates (northern Central African forest). The second reflects climate seasonality, contrasting equatorial areas with low water deficit and areas with high water deficit. The third component is correlated with maximum temperature peaks.

Although it would be easy to apply selection or regularization techniques to any zero-inflated binary tree Pólya-splitting (Z-TPS) model (see Section 3.2), some alternative modeling strategies such as multivariate Poisson log-normal (MPLN), multinomial, and Dirichlet multinomial (DM) models do not allow for this. There are multiple choices for binary partition trees (see Section 5). In this paper, the choice of the phylogenetic tree was driven by two ecological questions: (1) Does species inherit sensitivity or robustness to environmental factors along the species differentiation process? (2) Is the species scale the most appropriate level to estimate climate impacts on community distributions?

To sum-up data $i = 1, \ldots, 1571$, correspond to a triplet $\left(\boldsymbol{y}_i, \boldsymbol{X}_{(i)}, \tau\right), i = 1, \ldots, 1571$ quadrates, $\boldsymbol{y}_i = (y_{i1}, \ldots, y_{iJ})$, $J = 180$ the vector of the species abundances at the quadrat $i$, $\boldsymbol{X}'_{(i)}$ the vector of length five including the intercept, the three PCA components and $\tau$ the phylogenetic tree. Additionally, at each quadrat, the exact number of 0.5-ha sampled plots is known and used as an offset accounting for sampling efforts.

### 4.2 *Model Comparisons and criteria*

Does the use of a phylogenetic tree and multivariate zero-inflation improve goodness-of-fit (GoF) and prediction? Are PS distributions sufficiently flexible compared to alternatives such as the MPLN and its extensions?

To address the first question, we fitted several PS models using the negative binomial (NB) distribution as *global abundance distribution* for each: (i) the DM-NB model which uses the Dirichlet multinomial (DM) distribution as *split distribution*. This PS model ignores tree structure and zero-inflation. (ii) The GDM-NB models which use the generalized Dirichlet multinomial (GDM) distribution as *split distribution*. We considered two cases for the GDM-NB models; the case where the species are classified according to their abundance ($GDM^a$-NB) and the case where they are classified according to their phylogenetic order ($GDM^\phi$-NB). Since GDM distributions are based on cascading tree structure (Caplat et al., 2008), $GDM^a$-NB and $GDM^\phi$-NB are TPS models based on cascading trees structures. They do not include phylogenetic information and zero inflation. (iii) The zero-inflated versions of $GDM^a$-NB and $GDM^\phi$-NB models which are two Z-TPS models including zero inflation but not phylogenetic information. (iv) The DTM-NB model, which is a TPS model incorporating phylogenetic structure but not multivariate zero-inflation. (v) The zero-inflated version of the DTM-NB model, which is a Z-TPS model that include both zero-inflation and phylogenetic information.

To address the second question, model flexibility was further assessed using multivariate Poisson log-normal (MPLN) regression model and its zero-inflated version, where the zero-inflation parameters are regressed for each species (Chiquet et al., 2021; Batardière et al., 2024b). We consider unconstrained and constrained pseudo-residual covariance matrices for each case. Finally, the negative binomial model, the Poisson model and their zero-inflated versions are used as benchmarks.

Regression model performance was assessed using goodness-of-fit (GoF) criteria and mean absolute prediction errors on log-transformed data as accuracy measures. We applied a $k$-fold spatial cross-validation approach based on hierarchical clustering of the first three principal components (Dormann, 2007; Ploton et al., 2021). The number of clusters/folds ($k = 9$) was selected to balance within-cluster homogeneity and between-cluster separation (Hartigan, 1975). This strategy was designed to evaluate prediction robustness to environmental variation while ensuring a degree of independence between calibration and test datasets. Although the clusters were generally well ordered, cluster nine emerged as an outlier, showing higher averages for the first two components and an unexpectedly low value for the third component (Fig. 2).

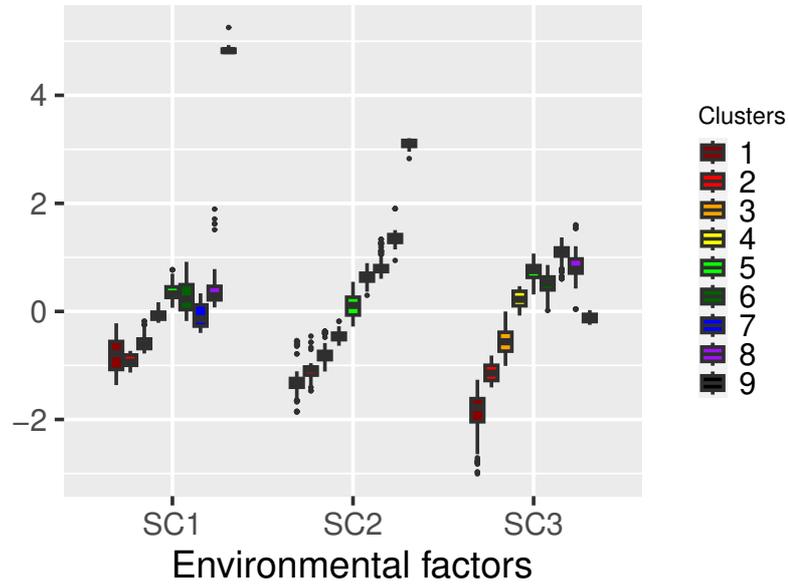

**Figure 2**: Boxplots of the first three principal components across the nine clusters.

The mean absolute error (MAE) on log1p-transformed predictions for each fold, $k = 1, \ldots, 9$, was defined as

$$\text{MAE}_{\log 1p}^{(k)} = \frac{1}{Jn_k} \sum_{i=1}^{n_k} \sum_{j=1}^{J} |\log(1 + y_{ij}) - \log(1 + \hat{y}_{ij})|$$

where $n_k$ is the number of observations in cluster $k$, and used as a measure of prediction accuracy. The $\text{MAE}_{\log 1p}$ metric balances sensitivity to low and high values, handles zeros naturally, and provides a robust error measure for skewed count data (Serhiyenko et al., 2016; Baron et al., 2021). All analyses were performed using adapted R packages (R Core Team, 2021), primarily *gamlss* (Rigby and Stasinopoulos, 2005) for model fitting and *ggplot2* (Wickham, 2016) for visualization.

### 4.3 *Results*

Model comparison reveals complex trade-offs between goodness-of-fit and predictive accuracy. Note that log-likelihoods for PLN models are approximate values corresponding to the *Evidence Lower Bound* (ELBO), obtained via variational expectation-maximization (Chiquet et al., 2019b) with $10^5$ maximum iterations and a $10^{-6}$ convergence threshold.

Based on penalized goodness-of-fit criteria (BIC and AIC), the multivariate PLN regression model with an unstructured covariance structure (MPLN) achieved the lowest BIC (723,000), despite its very large parameter space ($> 17,000$ parameters; Table 2). Introducing species-specific zero-inflation in the Z-MPLN model adds 720 additional parameters but results in a higher BIC (725,000). This increase reflects the strong penalization imposed by the BIC when the increase in model complexity is not compensated by a sufficient gain in log-likelihood. This can be explained by the fact that the latent Gaussian structure of the PLN model already captures a substantial part of the over-dispersion and excess zeros, leaving little residual information for a separate zero-inflation component. Similarly, the graphical lasso penalization approach applied within the MPLN framework did not yield a satisfactory trade-off between goodness-of-fit and model complexity, with BIC values exceeding 800,000. Consequently, MPLN-network models were not retained for further analysis.

Non-zero-inflated tree-based Pólya-splitting (TPS) models exhibited higher BIC values than MPLN models. However, incorporating zero-inflation dramatically improved their performance: with only around 600 additional parameters—negligible compared to MPLN's 17,000—zero-inflated TPS models (Z-DTM-NB, Z-GDM-NB) achieved BIC values close to MPLN (around 724,000–735,000) and outperformed Z-MPLN. This strong effect of zero-inflation, unlike what was observed with MPLN, unfortunately could not be fully attributed to the underlying tree structure: BIC remained comparable when species were assumed to be independent and distributed as zero-inflated NB. Both NB and PLN are mixture distribution models with gamma and log-Gaussian mixing distributions, respectively, exhibiting different distributional behaviors regarding extreme events (Valiquette et al., 2023). This reinforces the importance of choosing appropriate mixing distribution to model over-dispersed count data.

| Model(zi) | Param. | $(\frac{1}{2}\text{BIC})/1000$ | MAE / RMSE | | | |
|---|---|---|---|---|---|---|
| | | | $C_1^{211}$ | $C_2^{93}$ | $C_6^{296}$ | $C_9^{47}$ |
| DTM-NB | 1437(2025) | 735(727) | **1.19(1.16)/66(71)** | 1.05(0.99)/36(35) | 1.14(1.12)/60(60) | **1.14(1.06)/92(137)** |
| GDM$^0$-NB | 1437(2029) | 733(724) | 1.68(1.65)/214(251) | 1.54(1.53)/114(115) | 1.63(1.61)/175(188) | 1.31(1.30)/108(144) |
| GDM$^a$-NB | 1437(2069) | 733(724) | 1.77(1.71)/263(302) | 1.64(1.60)/122(127) | 1.72(1.71)/177(169) | 1.14(1.17)/127(147) |
| DM-NB | 725 | 760 | 1.57/65 | 1.45/45 | 1.22/67 | $8.04/9\times10^8$ |
| $\otimes$PO | 720(1440) | 4,272(3,094) | 1.32(1.33)/135(550) | **0.94(0.92)/21(20)** | **1.03(1.03)/38(38)** | $2.3(1.8)/4.43\times10^7(1.5\times10^6)$ |
| $\otimes$NB | 900(1620) | 740(725) | $1.48(1.35)/920(1.1\times10^5)$ | 1.0(0.93)/96(21) | 1.04(1.05)/38(39) | $3.6(1.86)/6.3\times10^{12}(3.9\times10^9)$ |
| $\otimes$PLN | 900(1620) | 755(738) | 1.84(-)/23681(2919) | 1.15(-)/590(-) | 1.18(-)/560(-) | $3.65(-)/2.42\times10^{10}(-)$ |
| MPLN | 17,010(17,730) | 723(725) | $2.7(-)/2.6\times10^6(-)$ | $2.25(-)/2.7\times10^5(-)$ | $2.08(-)/1.34\times10^4(-)$ | $3.84(-)/4.9\times10^9(-)$ |

**Table 2:** Model selection criteria (BIC) and predictive performance (MAE/RMSE) for spatial cross-validation. Values shown for clusters $C_1$ (211 sites), $C_2$ (93 sites), $C_6$ (296 sites), and $C_9$ (47 sites). Bold values indicate best performance per cluster. $\otimes$ denotes independent models. Values in parentheses correspond to zero-inflated model versions.

While BIC favored MPLN and zero-inflated models showed comparable fit, predictive performance evaluated through 9-fold spatial cross-validation revealed markedly different patterns (cf. Table 2). When assessed by mean absolute error (MAE) and root mean squared error (RMSE) averaged across all spatial clusters, the zero-inflated phylogenetic splitting model (Z-DTM-NB) exhibited the lowest overall prediction error (Cf. the full results of predictions in Appendix B). However, cluster-level analysis revealed substantial spatial heterogeneity in model performance: in 7 out of 9 cases, independent models achieved the lowest MAE and RMSE within individual clusters. For instance, in cluster 2, independent Poisson models achieved MAE=0.94 and RMSE=21, while in cluster 6, independent models ($\otimes$PO and $\otimes$NB) yielded MAE≈1.03–1.04 and RMSE≈38. In stark contrast, the joint multivariate Poisson log-normal model (MPLN)—despite its superior BIC—systematically underperformed in prediction with MAE values 2–3 times higher (e.g., MAE=2.25 for cluster 2, MAE=2.08 for cluster 6) and substantially inflated RMSE values. For the 9th cluster, characterized by clearly different environmental conditions, the best performance was achieved by the tree-structured negative binomial model (DTM-NB: MAE=1.14, RMSE=92), while all independent models collapsed with MAE>2.3 and RMSE> $10^7$.

Unlike the probit case where joint and independent models yield equivalent marginal predictions (Poggiato et al., 2021), Poisson log-normal models exhibit substantive differences because variance directly enters the expectation function ($\mathbb{E}[Z] = \exp(\mu + \sigma^2/2)$). In joint models, residual variance is decomposed into species-specific ($\sigma_j^2$) and shared covariance ($\sigma_{jk}$) components, which changes variance estimates and induces corresponding adjustments in mean parameters. As noted by Poggiato et al. (2021), residual covariances often reflect missing environmental covariates rather than true biotic interactions, leading to spurious associations that degrade marginal predictions—as evidenced by MPLN's poor predictive performance across all clusters despite excellent BIC. Tree-structured models (DTM-NB, GDM-NB) achieved performance comparable to independent models across most clusters (e.g., DTM-NB: MAE=1.19 for cluster 1, MAE=1.05 for cluster 2) while outperforming both approaches in the environmentally distinct 9th cluster, suggesting that their structured decomposition of species dependencies better captures some meaningful ecological relationships.

The addition of a zero-inflation component reveals substantial differences across model architectures. For PLN models (joint and independent), zero-inflation poses convergence issues and systematically degrades performance, as evidenced by Z-MPLN's worse BIC and the missing prediction values for Z-⊗PLN in Table 2. Independent models (⊗PO, ⊗NB) exhibit strong instability: while zero-inflation can improve MAE in some clusters (cluster 2: ⊗NB from MAE=1.0 to 0.93, $-7\%$), RMSE reveals catastrophic errors (cluster 1: RMSE from 920 to $1.1\times10^5$, $\times120$ increase). This MAE/RMSE divergence indicates that these models predict correctly on average but commit extreme errors particularly for zero-abundant observations, suggesting misspecification of the zero-generating process. In contrast, tree-structured models show consistent behavior: zero-inflation systematically improves MAE (cluster 9: $1.14\rightarrow1.06$, $-7\%$; cluster 2: $1.05\rightarrow0.99$, $-6\%$) while keeping RMSE variations moderate ($\pm25\%$ maximum vs $\times100+$ for independent models). This coherence between both metrics suggests that the hierarchical structure captures structural zeros related to shared phylogenetic traits, leaving only residual zeros for the zero-inflation component to model.

In summary, incorporating species dependencies through tree structure substantially improves predictive performance while maintaining competitive goodness-of-fit, with tree-structured models demonstrating superior robustness across environmentally diverse clusters. Critically, model complexity does not guarantee better predictions: simple models (DM-NB, 725 parameters) matched or exceeded complex approaches (MPLN, 17,000 parameters) in cross-validation despite worse BIC, while phylogenetic structure consistently provided the most reliable framework. The zero-inflation component proved most effective when combined with hierarchical structures, improving predictions particularly for range-restricted species where environmental filtering primarily governs presence/absence rather than local abundances. These findings underscore that model architecture—specifically the integration of phylogenetic relationships—matters more than parameter count or in-sample fit for accurate community-level predictions.

### 4.4 *Ecological interpretations*

The use of the phylogenetic tree enabled a sequential and ecologically meaningful interpretation of the zero-inflated DTM-NB model outputs. At each node, marginal abundances reflected the sensitivity of species group abundances to global environmental changes, while the split distributions indicated how such changes might influence differentiation processes.

At the root level, total abundance was primarily and negatively affected by the third component (SC3), which is directly associated with temperature peaks during the dry season. The other two covariates (SC1 and SC2) showed no significant effects. This result highlights that global warming constitutes one of the main climatic drivers shaping overall floral abundance in Central African rainforests, with a detrimental impact (Wright, 2010).

Except for the root node representing total abundance, we employed the *Relative Size Effect* (RSE) to quantify the influence of external factors on marginal abundances along the tree. RSE is defined as the ratio between the size effect of a covariate on node abundance and the predicted abundance at that node. It thus represents the proportionate change in abundance associated with a one-unit increase in an environmental variable. The results of the RSE analysis across the phylogenetic tree, for the component related to temperature peaks, are presented in Figure 3. Overall, this figure shows that the colors observed at the genera level are inherited from those at deeper nodes of the phylogenetic tree, suggesting heritable traits influencing tree species' responses to global warming (Martins et al., 2019). Moreover, the results make it possible to visualize clusters of genera sharing similar colors, indicating that genetic relatedness is a key factor shaping how tree species respond to temperature extremes. Three main categories of genera can be distinguished: (i) genera highly threatened by global warming (dark red), such as Aucoumea, Cola, and Ceiba; (ii) genera that appear favored by warming (dark blue), such as Beilschmiedia, Elaeis, and Tabernaemontana; and (iii) genera that are largely unaffected (gray), such as Panda, Polyalthia, and Mansonia. Taken together, these findings illustrate how temperature peaks modulate the biodiversity structure of Central African rainforests, identifying species likely to decline or thrive depending on the severity of warming, and revealing a clear link between responses to global warming and phylogenetic history.

Conversely, the zero-inflated DTM-NB model provides insights into how climatic factors shape species differentiation. Focusing on the phylogenetic sub-tree dated at approximately 180 million years ago, mainly composed of four major clades (Asterids, Olacaceae, Rosids, and Magnoliales), the results highlight how specific climatic covariates drive divergence between pairs of genera or groups of genera. For instance, in the differentiation between Asterids and Olacaceae, Asterids are favored by covariates related to evapotranspiration and temperature peaks, whereas Olacaceae are primarily associated with seasonality. In contrast, in the comparison between Elaeis and Dracaena, all climatic covariates favor Dracaena. Finally, the model sheds light on how climatic gradients contribute to internal differentiation processes within a lineage. Along the differentiation pathway leading to Asterids, when effects are statistically significant, the evapotranspiration-related component (SC1) and the temperature-peak component (SC3) favor the Asterids lineage at each successive split, whereas the seasonality component (SC2) disfavors this lineage throughout the pathway.

## 5. Discussion

The objective of this paper was to develop a flexible approach for modeling multivariate count data from highly bio-diverse ecosystem. We proposed a new class of *tree Pólya splitting* distributions specifically designed to account for joint zero inflation.

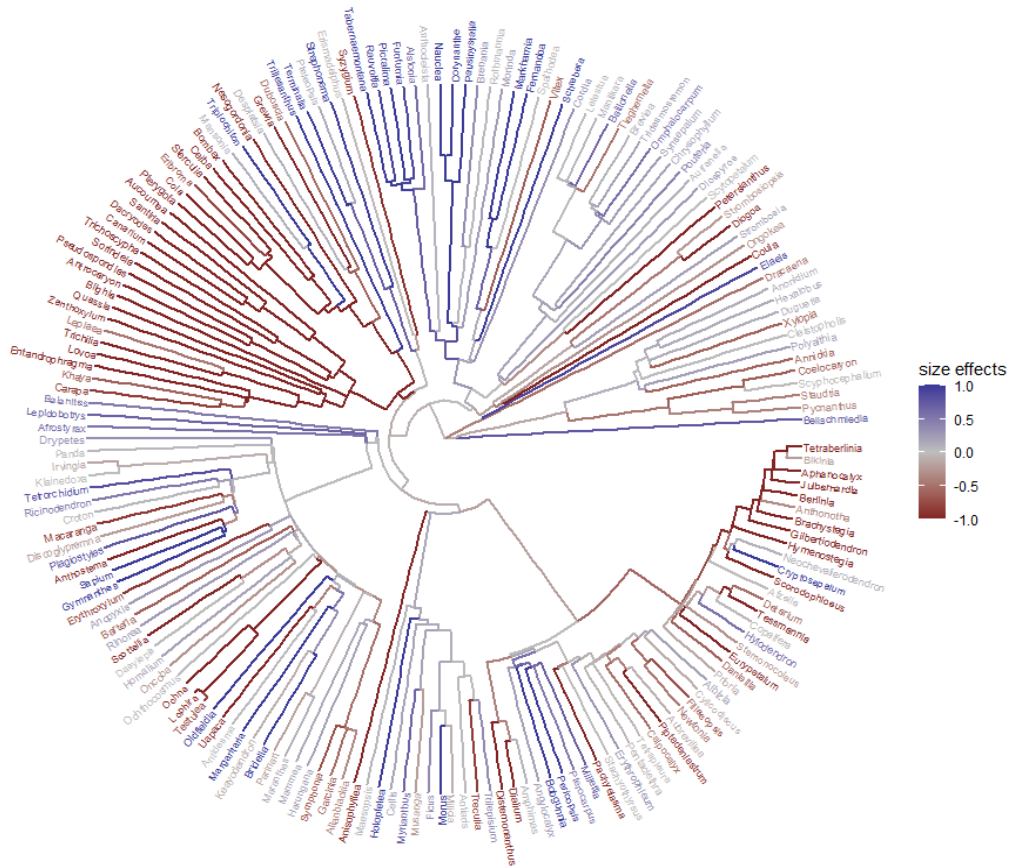

**Figure 3**: Relative size effects (sizes effects divided by predicted abundance) of the component related to temperatures peaks following the phylogenetic tree. In grey: values close to zero; In blue: value close to 1; In red: values close to -1.

This model has been applied for modeling tree genera count data from tropical rainforest, taking into account phylogenetic dependencies between the tree genera.

From a statistical perspective, the results highlight the need to balance statistical flexibility, computational efficiency, and biological interpretability when modeling species communities. Although the multivariate Poisson log-normal (MPLN) model achieved the lowest Bayesian Information Criterion (BIC), its weak predictive performance reveals the classical trade-off between goodness-of-fit and predictive generalization (Kp, 1998). The decline in predictive accuracy with increasing parameter dimensionality (>17,000 parameters) further suggests that the MPLN overfits residual covariances, which likely capture unmeasured environmental co-variables rather than real biotic dependencies (Poggiato et al., 2021). This behavior exemplifies how hierarchical Bayesian-type models, though powerful for joint species modeling, can blur the boundary between ecological signal and statistical artifact when residual structures compensate for missing co-variates. Tree-structured models, by contrast, provide a more parsimonious yet biologically meaningful representation of species co-variation. Despite having over ten times fewer parameters than the MPLN, they achieved comparable—or even better—BIC values and, more importantly, displayed much greater predictive robustness across spatial clusters. The inclusion of a zero-inflation component (Z-DTM-NB) further improved both Mean Absolute Error (MAE) and Root Mean Square Error (RMSE) without exhibiting the instability observed in independent models (⊗PO, ⊗NB and ⊗PLN). This result suggests that the hierarchical phylogenetic structure acts as a regularization mechanism, partitioning ecological variability along evolutionary axes and effectively distinguishing structural zeros (species absences due to habitat or lineage constraints) from sampling zeros (absences due to stochastic or sampling effects). Hence, incorporating zero inflation within the DTM-NB framework complements—rather than obscures—the phylogenetic structure during modeling.

From an ecological perspective, the phylogenetic decomposition provides an integrated interpretation of climatic effects on both abundance and differentiation. The strong negative association between total abundance and temperature peaks (SC3) confirms that rising thermal extremes exert a detrimental effect on the persistence of Central African forest flora, consistent with previous studies in tropical systems (Wright et al., 2009). Furthermore, the analysis of Relative Size Effects (RSE) along the phylogenetic tree reveals that closely related genera exhibit similar responses to climatic stress, suggesting partial heritability of traits related to drought or heat tolerance (Cavender-Bares et al., 2009). This pattern allows the distinction between genera vulnerable to warming (e.g., *Aucoumea, Cola, Ceiba*) and those potentially favored by such conditions (e.g., *Beilschmiedia, Elaeis, Tabernaemontana*), illustrating divergent adaptive trajectories embedded within phylogenetic history. The differentiation analysis of the sub-tree dating back approximately 180 million years further examines how climatic covariates are associated with floristic divergence among major clades such as Asterids, Olacaceae, Rosids, and Magnoliales. In addition, examining climatic covariates along the differentiation pathway leading to Asterids shows that, when statistically significant, evapotranspiration and temperature peaks tend to favor the Asterids lineage at successive splits, whereas seasonality acts in the opposite direction along this pathway. Together, these patterns suggest that phylogenetic relatedness constrains species' climatic associations, as lineages sharing similar functional traits display more homogeneous responses to climatic gradients, potentially limiting adaptive diversity and increasing vulnerability to ongoing warming.

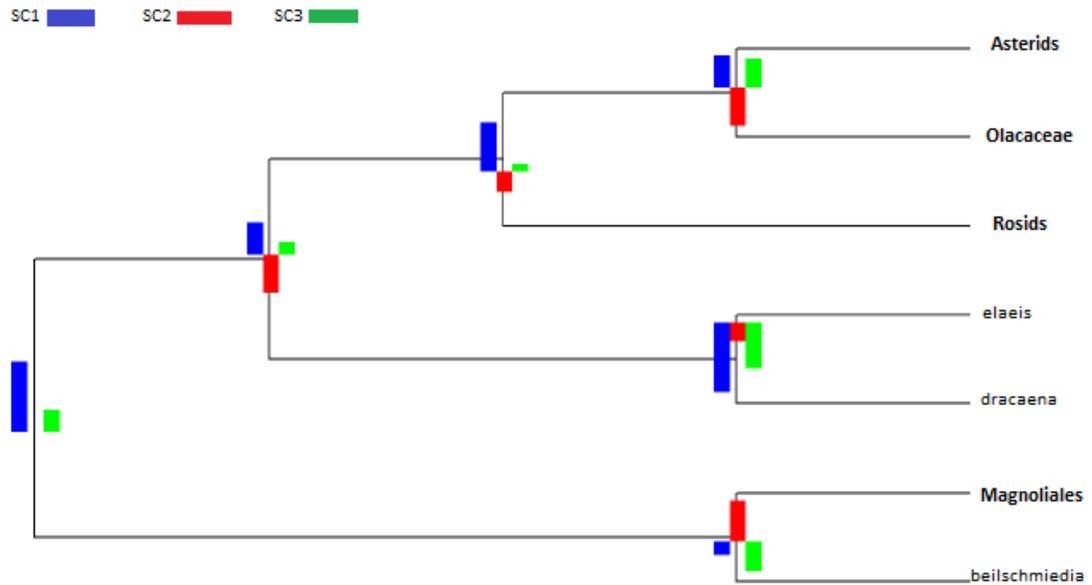

**Figure 4**: Effect of variables on tree species differentiation. The colors represent the variables (components): blue for the first component (SC1), red for the seond component (SC2), green for the third component (SC3). At each node, the variables represented are those with a significant effect. Sizes of bars are proportional to the intensities of the regression coefficients and oriented in the direction where the variable has a positive effect.

Overall, the results demonstrate that, in joint species distribution modeling, model architecture matters more than parameter richness. Structured and parsimonious models grounded in phylogenetic relationships yield more reliable predictions and more interpretable ecological inferences than high-dimensional unstructured multivariate approaches. The Z-TPS framework therefore emerges as a promising compromise between statistical rigor, computational simplicity, and ecological realism. Nevertheless, the modeling architecture proposed here can still be optimized. Given that Z-TPS performance depends on the underlying tree, one potential extension would be to explore the tree space to identify an optimal structure. Additionally, more general graphical structures than rooted trees may better represent inter-specific dependencies. Developing a Pólya-splitting–type models for such generalized graph structures, while maintaining computational simplicity, would be interesting for future research.

## References

Aitchison, J. and Ho, C. (1989). The multivariate poisson-log normal distribution. *Biometrika* **76,** 643–653.

Baron, M., Vermeirssen, V., De Tender, C., De Schaetzen, V., and Dawyndt, P. (2021). Comparison of normalization methods for high-throughput rna sequencing data: A review. *Frontiers in Genetics* **12,** 627836.

Bartholomew, D., Knott, M., and Moustaki, I. (2011). *Latent Variable Models and Factor Analysis: A Unified Approach.*

Batardière, B., Chiquet, J., Gindraud, F., and Mariadassou, M. (2024a). Zero-inflation in the multivariate poisson lognormal family.

Batardière, B., Chiquet, J., Gindraud, F., and Mariadassou, M. (2024b). Zero-inflation in the multivariate poisson lognormal family. *arXiv preprint* .

Caplat, P., Anand, M., and Bauch, C. (2008). Interactions between climate change, competition, dispersal, and disturbances in a tree migration model. *Theoretical Ecology* **1,** 209–220.

Cavender-Bares, J., Kozak, K. H., Fine, P. V., and Kembel, S. W. (2009). The merging of community ecology and phylogenetic biology. *Ecology letters* **12,** 693–715.

Chen, J. and Li, H. (2013). Variable selection for sparse dirichlet-multinomial regression with an application to microbiome data analysis. *The annals of applied statistics* **7,**.

Chib, S. and Greenberg, E. (1998). Analysis of multivariate probit models. *Biometrika* **85,** 347–361.

Chiquet, J., Mariadassou, M., and Robin, S. (2021). The poisson-lognormal model as a versatile framework for the joint analysis of species abundances. *Frontiers in Ecology and Evolution* **9,** 588292.

Chiquet, J., Robin, S., and Mariadassou, M. (2019a). Variational inference for sparse network reconstruction from count data. In Chaudhuri, K. and Salakhutdinov, R., editors, *Proceedings of the 36th International Conference on Machine Learning*, volume 97 of *Proceedings of Machine Learning Research*, pages 1162–1171.

Chiquet, J., Robin, S., and Mariadassou, M. (2019b). Variational inference for sparse network reconstruction from count data. In *International Conference on Machine Learning*, pages 1162–1171. PMLR.

Clément, J. and Vieilledent, G. (2023). *jSDM: Joint Species Distribution Models*. R package version 0.2.6.

Dormann, C. F. (2007). Effects of incorporating spatial autocorrelation into the analysis of species distribution data. *Global Ecology and Biogeography* **16,** 129–138.

Dyczka, W. (1973). On the multidimensional polya distribution. *Commentationes Mathematicae* **17,**.

Elith, J. and Leathwick, J. R. (2009). Species distribution models: ecological explanation and prediction across space and time. *Annual review of ecology, evolution, and systematics* **40,** 677–697.

Friedman, J., Hastie, T., and Tibshirani, R. (2007). Sparse inverse covariance estimation with the graphical lasso. *Biostatistics* **9,** 432–441.

Gentry, A. H. (1988). Changes in plant community diversity and floristic composition on environmental and geographical gradients. *Annals of the Missouri botanical garden* pages 1–34.

Gibaud, J., Bry, X., and Trottier, C. (2025). Generalized linear model based on latent factors and supervised components. working paper or preprint.

Gross, J. L., Yellen, J., and Anderson, M. (2018). *Graph theory and its applications*. Chapman and Hall/CRC.

Haegeman, B. and Etienne, R. (2008). Relaxing the zero-sum assumption in neutral biodiversity theory. *Journal of Theoretical Biology* **252,** 288–294.

Haegeman, B. and Etienne, R. S. (2017). A general sampling formula for community structure data. *Methods in Ecology and Evolution* **8,** 1506–1519.

Hartigan, J. (1975). *Clustering Algorithms*. Out-of-print Books on demand. Wiley.

Hubbell, S. (2001). A unified theory of biodiversity and biogeography–princeton university press. *Princeton, NJ* .

Inouye, D. I., Yang, E., Allen, G. I., and Ravikumar, P. (2017). A review of multivariate distributions for count data derived from the poisson distribution. *WIREs Computational Statistics* **9,** e1398.

Ives, A. and Helmus, M. (2011). Generalized linear mixed models for phylogenetic analyses of community structure. *Ecological Monograph* **81,** 511–525.

Jones, M. C. and Marchand, (2019). Multivariate discrete distributions via sums and shares. *Journal of Multivariate Analysis* **171, 83-93.,** 83–93.

Kp, B. (1998). Model selection and multimodel inference. *A practical information-theoretic approach* .

Manor, O. and Borenstein, E. (2015). Musicc: a marker genes based framework for metagenomic normalization and accurate profiling of gene abundances in the microbiome. *Genome Biology* **16,**.

Martins, F., Kruuk, L., Llewelyn, J., Moritz, C., and Phillips, B. (2019). Heritability of climate-relevant traits in a rainforest skink. *Heredity* **122(1),** 41–52.

O'Brien, S. and Dunson, D. (2004). Bayesian multivariate logistic regression. *Biometrics* **60,** 739–746.

Ovaskainen, O., Hottola, J., and Siitonen, J. (2010). Modeling species co-occurrence by multivariate logistic regression generates new hypotheses on fungal interactions. *Ecology* **91,** 2514–2521.

Ovaskainen, O., Tikhonov, G., Norberg, A., Guillaume Blanchet, F., Duan, L., Dunson, D., Roslin, T., and Abrego, N. (2017). How to make more out of community data? a conceptual framework and its implementation as models and software. *Ecology letters* **20,** 561–576.

Peyhardi, J. and Fernique, P. (2017). Characterization of convolution splitting graphical models. *Statistics & Probability Letters* **126,** 59–64.

Peyhardi, J., Fernique, P., and Durand, J.-B. (2021). Splitting models for multivariate count data. *Journal of Multivariate Analysis* **181,** 104677.

Peyhardi, J., Laroche, F., and Mortier, F. (2024). Pólya-splitting distributions as stationary solutions of multivariate birth–death processes under extended neutral theory. *Journal of Theoretical Biology* page 111755.

Pichler, M. and Hartig, F. (2021). A new joint species distribution model for faster and more accurate inference of species associations from big community data. *Methods in Ecology and Evolution* **12,** 2159–2173.

Ploton, P., Mortier, F., and Réjou-Méchain, M. e. a. (2021). Spatial validation reveals poor predictive performance of large-scale ecological mapping models. *Nature Communications* **11,** 5440.

Poggiato, G., Münkemüller, T., Bystrova, D., Arbel, J., Clark, J. S., and Thuiller, W. (2021). On the interpretations of joint modeling in community ecology. *Trends in Ecology & Evolution* **36,** 391–401.

Pollock, L. and et al. (2014). Understanding co-occurrence by modelling species simultaneously with a joint species distribution model (jsdm). *Methods in Ecology and Evolution* **5,** 397–406.

R Core Team (2021). *R: A Language and Environment for Statistical Computing.* R Foundation for Statistical Computing, Vienna, Austria.

Réjou-Méchain, M., Mortier, F., Bastin, J.-F., Cornu, G., Barbier, N., Bayol, N., Bénédet, F., Bry, X., Dauby, G., Deblauwe, V., et al. (2021). Unveiling african rainforest composition and vulnerability to global change. *Nature* **593,** 90–94.

Rigby, R. A. and Stasinopoulos, D. M. (2005). Generalized additive models for location, scale and shape. *Journal of the Royal Statistical Society Series C: Applied Statistics* **54,** 507–554.

Rigby, R. A., Stasinopoulos, M. D., Heller, G. Z., and De Bastiani, F. (2019). *Distributions for modeling location, scale, and shape: Using GAMLSS in R.* Chapman and Hall/CRC.

Serhiyenko, V., Mamun, S., Ivan, J., and Ravishanker, N. (2016). Fast bayesian inference for modeling multivariate crash counts. *Analytic Methods in Accident Research* **9,** 44–53.

Tang, Z.-Z. and Chen, G. (2019). Zero-inflated generalized dirichlet multinomial regression model for microbiome compositional data analysis. *Biostatistics* **20,** 698–713.

Tikhonov, G., Ovaskainen, O., Oksanen, J., De Jonge, M., Opedal, O., and Dallas, T. (2022). *Hmsc: Hierarchical Model of Species Communities.* R package version 3.0-13.

Valiquette, S., Marchand, É., Peyhardi, J., Toulemonde, G., and Mortier, F. (2024). Tree p {\'o} lya splitting distributions for multivariate count data. *arXiv preprint arXiv:2404.19528* .

Valiquette, S., Toulemonde, G., Peyhardi, J., Marchand, É., and Mortier, F. (2023). Asymptotic tail properties of poisson mixture distributions. *Stat* **12,** e622.

Wang, T. and Zhao, H. (2017). A dirichlet-tree multinomial regression model for associating dietary nutrients with gut microorganisms. *Biometrics* **73,** 792–801.

Warton, D., Blanchet, F., O'Hara, R., Ovaskainen, O., Taskinen, S., Walker, S., and Hui, F. (2015). So many variables: Joint modeling in community ecology. *Trends in Ecology and Evolution* **30,** 766–779.

Warton, D. I., Foster, S. D., De'ath, G., Stoklosa, J., and Dunstan, P. K. (2015). Model-based thinking for community ecology. *Plant Ecology* **216,** 669–682.

Wickham, H. (2016). *ggplot2: Elegant Graphics for Data Analysis.* Springer-Verlag New York.

Wright, S. J. (2010). The future of tropical forests. *Annals of the New York Academy of Sciences* **1195,** 1–27.

Wright, S. J., MULLER-LANDAU, H. C., and Schipper, J. (2009). The future of tropical species on a warmer planet. *Conservation biology* **23,** 1418–1426.

Zhang, P., Pitt, D., and Wu, X. (2022). A comparative analysis of several multivariate zero-inflated and zero-modified models with applications in insurance.

## Appendix A : characteristics of the Z-TPS distribution

### A1: Factorial moments

Let $X$ be a random variable and $B$ a node of the tree $\mathcal{I}$ different to its root node $\Omega$. Let's denote $(X)_k = X(X-1)\ldots(X-k+1)$ and $\mathcal{A}_B$ the set of ancestor of $B$ except the root and including $B$. Let $\boldsymbol{Y} = \{Y_j\}_{j\in\Omega}$ a random count vector following the zero-inflated tree Pólya-splitting distribution based on the tree $\mathcal{I}$. Let's demonstrate by recurrence that:

$$\mu_k\left[(|\boldsymbol{Y}_B|)_k\right] = \left\{\prod_{D\in\mathcal{A}_B}\tilde{p}_D^{(k)}\right\}\times\mu_k(\Omega).$$

- If $B$ has one ancestor, given $|\boldsymbol{Y}| = n$, $|\boldsymbol{Y}_B| \sim \pi_B\delta_0 + \pi_{\bar{B}}\delta_n + (1 - \pi_B - \pi_{\bar{B}})\mathcal{P}_n^{[c^B]}(\theta_B, \theta_{\bar{B}})$. Since $\mathrm{E}\left[(n)_k\right] = \mu_k(\Omega)$,

$$\mu_k\left[(|\boldsymbol{Y}_B|)_k\right] = \mathrm{E}\left[\mathrm{E}\left[(|\boldsymbol{Y}_B|)_k/n\right]\right] = \tilde{p}_B^{(k)}\mu_k(\Omega).$$

- If $B$ has two ancestors, by denoting $F$ his father, given $|\boldsymbol{Y}_F| = n$, $|\boldsymbol{Y}_B| \sim \pi_B\delta_0 + \pi_{\bar{B}}\delta_n + (1 - \pi_B - \pi_{\bar{B}})\mathcal{P}_n^{[c^B]}(\theta_B, \theta_{\bar{B}})$. Since $\mathrm{E}\left[(n)_k\right] = \mathrm{E}\left[(|\boldsymbol{Y}_F|)_k\right] = \tilde{p}_F^{(k)}\mu_k(\Omega)$,

$$\mu_k\left[(|\boldsymbol{Y}_B|)_k\right] = \mathrm{E}\left[\mathrm{E}\left[(|\boldsymbol{Y}_B|)_k/n\right]\right] = \tilde{p}_B^{(k)}\tilde{p}_F^{(k)}\mu_k(\Omega).$$

- We supposed the result true for any node with $p-1$ ancestors ($p \geqslant 2$). For any node $B$ with $p$ ancestors, by denoting $F$ the

father of $B$, given $|\boldsymbol{Y}_F| = n$, $|\boldsymbol{Y}_B| \sim \pi_B \delta_0 + \pi_{\bar{B}} \delta_n + (1 - \pi_B - \pi_{\bar{B}}) \mathcal{P}_n^{[c^B]}(\theta_B, \theta_{\bar{B}})$. Since $\mathrm{E}\left[(n)_k\right] = \left\{ \prod_{D \in \mathcal{A}_F} \tilde{p}_D^{(k)} \right\} \times \mu_k(\Omega)$,

$$\mu_k\big[(|\boldsymbol{Y}_B|)_k\big] = \mathrm{E}\left[\mathrm{E}\big[(|\boldsymbol{Y}_B|)_k/n\big]\right] = \tilde{p}_B^{(k)} \left\{ \prod_{D \in \mathcal{A}_F} \tilde{p}_D^{(k)} \right\} \times \mu_k(\Omega) = \left\{ \prod_{D \in \mathcal{A}_B} \tilde{p}_D^{(k)} \right\} \times \mu_k(\Omega) \blacksquare$$

Notice that, for $B = \{j\}$, we have $\mathcal{A}_B = \mathcal{A}_j$.

### A2: Marginals

The proof is similar to that of *Proposition 5* in Valiquette et al. (2024).

### A3: Covariances

For the general form of the covariance, the proof is similar to that of *Proposition 12* in Valiquette et al. (2024). Now , let us focus on covariance between two sibling nodes. Let $S_1$, $S_2$ be two sibling nodes, $S$ theirs father node and $|Y_S| = n$. We introduce a latent categorical variable $H \in \{0, 1, 2\}$ that determines the allocation regime of $n$ between $|Y_{S_1}|$ and $|Y_{S_2}|$:

$$P(H = 0) = \pi_{S_1}, \quad P(H = 1) = \pi_{S_2}, \quad P(H = 2) = 1 - \pi_{S_1} - \pi_{S_2}.$$

The pair $(|Y_{S_1}|, |Y_{S_2}|)$ can thus be expressed as

$$(|Y_{S_1}|, |Y_{S_2}|) = \mathbb{1}_{\{H=0\}}(0, n) + \mathbb{1}_{\{H=1\}}(n, 0) + \mathbb{1}_{\{H=2\}}(|Z_{S_1}|, |Z_{S_2}|),$$

where $(|Z_{S_1}|, |Z_{S_2}|) \sim P_n^{[c_S]}(\theta_{S_1}, \theta_{S_2})$. Hence, the joint law of $(|Y_{S_1}|, |Y_{S_2}|)$ can be written as

$$(|Y_{S_1}|, |Y_{S_2}|) \sim \pi_{S_1} \, \delta_{(0,n)} + \pi_{S_2} \, \delta_{(n,0)} + (1 - \pi_{S_1} - \pi_{S_2}) \, P_n^{[c-S]}(\theta_{S_1}, \theta_{S_2}).$$

Using the law of total covariance with respect to both $H$ and $n$:

$$\mathrm{Cov}(|Y_{S_1}|, |Y_{S_2}|) = E_n[E_H[\mathrm{Cov}(|Y_{S_1}|, |Y_{S_2}|/n, H)]] + E_n[\mathrm{Cov}_H(E[|Y_{S_1}|/n, H], E[|Y_{S_2}|/n, H])] + \mathrm{Cov}_n(E[|Y_{S_1}|/n], E[|Y_{S_2}|/n]).$$

By calculating the different terms of the covariance and by denoting $\pi_S = \pi_{S_1} + \pi_{S_2}$, we obtain:

$$\mathrm{Cov}(|\boldsymbol{Y}_{S_1}|, |\boldsymbol{Y}_{S_2}|) = (1 - \pi_S) \cdot p_{S_1} \cdot p_{S_2} \frac{\theta_{S_1} + \theta_{S_2}}{\theta_{S_1} + \theta_{S_2} + c_S} \mu_2(S) - \tilde{p}_{S_1} \cdot \tilde{p}_{S_2} \cdot \mu^2(S) \blacksquare$$

Notive that, When $\pi_1 = \pi_2 = 0$, the latent variable $H$ always takes value 2, corresponding to the standard non-inflated Pólya split given by (Valiquette et al., 2024):

$$\mathrm{Cov}(|\boldsymbol{Y}_{S_1}|, |\boldsymbol{Y}_{S_2}|) = p_{S_1} \cdot p_{S_2} \frac{\theta_{S_1} + \theta_{S_2}}{\theta_{S_1} + \theta_{S_2} + c_S} \mu_2(S) - p_{S_1} \cdot p_{S_2} \cdot \mu^2(S).$$

### Appendix B: Full prediction table

| Mod | $C_1^{211}$ | $C_2^{93}$ | $C_3^{290}$ | $C_4^{62}$ | $C_5^{296}$ | $C_6^{280}$ | $C_7^{217}$ | $C_8^{75}$ | $C_9^{47}$ | Global |
|---|---|---|---|---|---|---|---|---|---|---|
| DTM-NB | 1.19 | 1.05 | 1.36 | 1.20 | 1.25 | 1.14 | 1.10 | 0.89 | 1.14 | 1.19 |
| | **66** | 36 | 69 | 48 | 47 | 60 | 32 | 35 | **92** | **54** |
| Z-DTM-NB | **1.16** | 0.99 | 1.32 | 1.14 | 1.19 | 1.12 | 1.01 | **0.85** | **1.06** | **1.14** |
| | 71 | 35 | 67 | 48 | 47 | 60 | 31 | 35 | 137 | 55 |
| GDM$^\alpha$-NB | 1.68 | 1.54 | 1.63 | 1.68 | 1.53 | 1.63 | 1.22 | 1.17 | 1.31 | 1.52 |
| | 214 | 114 | 213 | 173 | 189 | 175 | 105 | 90 | 108 | 171 |
| Z-GDM$^\alpha$-NB | 1.65 | 1.53 | 1.65 | 1.68 | 1.53 | 1.61 | 1.23 | 1.18 | 1.3 | 1.52 |
| | 251 | 115 | 219 | 175 | 195 | 188 | 93 | 92 | 144 | 180 |
| GDM$^o$-NB | 1.77 | 1.64 | 1.78 | 1.81 | 1.69 | 1.72 | 1.24 | 1.19 | 1.14 | 1.62 |
| | 263 | 122 | 198 | 158 | 177 | 177 | 83 | 75 | 127 | 169 |
| Z-GDM$^o$-NB | 1.72 | 1.61 | 1.76 | 1.79 | 1.68 | 1.71 | 1.15 | 1.14 | 1.18 | 1.59 |
| | 302 | 127 | 200 | 162 | 172 | 170 | 116 | 102 | 147 | 179 |
| DM-NB | 1.57 | 1.45 | 1.53 | 1.52 | 1.07 | 1.22 | 1.05 | 0.99 | 8.04 | 1.5 |
| | 65 | 46 | 81 | 58 | 56 | 67 | 32 | 47 | $9 \times 10^8$ | $2.7 \times 10^7$ |
| $\otimes$PO | 1.32 | 0.94 | 1.27 | 1.14 | 1.23 | 1.03 | 1.22 | 0.89 | 2.3 | 1.21 |
| | 136 | **21** | 44 | 33 | 37 | 38 | 27 | 17 | $4 \times 10^7$ | $13 \times 10^5$ |
| $\otimes$Z-PO | 1.33 | **0.92** | 1.26 | 1.15 | 1.20 | **1.03** | 1.11 | 0.87 | 1.81 | 1.17 |
| | 553 | **21** | **43** | 33 | 37 | **38** | 23 | **16** | $1.5 \times 10^6$ | $4 \times 10^4$ |
| $\otimes$NB | 1.48 | 1.00 | **1.23** | **1.07** | **1.13** | 1.04 | 1.13 | 0.92 | 3.60 | 1.24 |
| | 915 | 86 | 45 | **31** | **35** | 38 | 21 | 29 | $6 \times 10^{12}$ | $18 \times 10^{10}$ |
| $\otimes$Z-NB | 1.35 | 0.94 | 1.37 | 1.13 | 1.15 | 1.05 | **0.99** | **0.85** | 1.86 | 1.17 |
| | $1.1 \times 10^5$ | **21** | 82 | 33 | 36 | 39 | **20** | 17 | $3.8 \times 10^9$ | $1.1 \times 10^7$ |
| $\otimes$PLN | 1.85 | 1.15 | 1.43 | 1.23 | 1.39 | 1.18 | 1.38 | 1.05 | 3.65 | 1.45 |
| | $2.3 \times 10^4$ | 589 | 727 | 168 | 284 | 562 | 452 | 302 | $2.4 \times 10^{10}$ | $7.2 \times 10^7$ |
| $\otimes$Z-PLN | - | - | - | - | - | - | - | - | - | - |
| | - | - | - | - | - | - | - | - | - | - |
| MPLN | 2.71 | 2.25 | 2.31 | 2.29 | 2.48 | 2.08 | 2.17 | 2.06 | 3.85 | 2.37 |
| | $2 \times 10^6$ | $2 \times 10^5$ | $3 \times 10^5$ | $\times 10^5$ | $2 \times 10^5$ | $10^4$ | $10^5$ | $10^6$ | $5 \times 10^9$ | $10^8$ |
| Z-MPLN | - | - | - | - | - | - | - | - | - | - |
| | - | - | - | - | - | - | - | - | - | - |

Table 1: Full prediction results. In columns the clusters, in line the models, in values the predictions results. For each model, the first sub-line represent the $MAE_{log1p}$ results and the second sub-line the RMSE.